\documentclass[aps,prc,preprint,showkeys,superscriptaddress,amsmath,amssymb,floatfix]{revtex4}

\usepackage[dvips]{graphicx,color}
\usepackage{dcolumn}
\usepackage{bm}

\def\pmb#1{\setbox0=\hbox{#1}%
  \kern-.025em\copy0\kern-\wd0 
  \kern.05em\copy0\kern-\wd0
  \kern-.025em\raise.0433em\box0 }

\newcommand{\Figuretable}[1]{%
  \begin{center} --------- {\bf #1} --------- \\ \end{center}} 
\def\lambdabar{\protect\@lambdabar}
\def\@lambdabar{%
\relax
\bgroup
\def\@tempa{\hbox{\raise.73\ht0
\hbox to0pt{\kern.25\wd0\vrule width.5\wd0
height.1pt depth.1pt\hss}\box0}}%
\mathchoice{\setbox0\hbox{$\displaystyle\lambda$}\@tempa}%
{\setbox0\hbox{$\textstyle\lambda$}\@tempa}%
{\setbox0\hbox{$\scriptstyle\lambda$}\@tempa}%
{\setbox0\hbox{$\scriptscriptstyle\lambda$}\@tempa}%
\egroup
}

\begin{document}

\preprint{}

\title{\boldmath
Production of doubly strange hypernuclei via $\Xi^-$ doorways 
in the $^{16}$O($K^-$,~$K^+$) reaction at 1.8 GeV/c
}

\author{Toru~Harada}%
\email{harada@isc.osakac.ac.jp}
\affiliation{%
Research Center for Physics and Mathematics,
Osaka Electro-Communication University, Neyagawa, Osaka, 572-8530, Japan
}

\author{Yoshiharu~Hirabayashi}%
\affiliation{%
Information Initiative Center, 
Hokkaido University, Sapporo, 060-0811, Japan
}

\author{Atsushi~Umeya}%
\affiliation{%
Nishina Center for Accelerator-Based Science, RIKEN, Wako 351-0198, Japan
}

\date{\today}

\begin{abstract}

We examine theoretically production of doubly strange hypernuclei, 
$^{~16}_{~\Xi^-}$C and $^{~16}_{\Lambda\Lambda}$C, in double-charge exchange 
$^{16}$O($K^-$,~$K^+$) reactions using a distorted-wave impulse approximation. 
The inclusive $K^+$ spectrum at the incident momentum $p_{K^-}=$ 1.8 GeV/c and  
scattering angle $\theta_{\rm lab}=0^\circ$ is estimated in a one-step 
mechanism, $K^-p \to K^+ \Xi^-$ via $\Xi^-$ doorways caused by a 
$\Xi^-p\text{--}\Lambda \Lambda$ coupling.
The calculated spectrum in the $\Xi^-$ bound region indicates 
that the integrated cross sections are on the order of $7\text{--}12$ nb/sr 
for significant $1^-$ excited states with 
${^{14}{\rm C}(0^+, 2^+)}\otimes s_{\Lambda}p_{\Lambda}$ configurations 
in $^{16}_{\Lambda\Lambda}$C via the doorway states of the spin-stretched 
${^{15}{\rm N}({1/2}^-, {3/2}^-)}\otimes s_{\Xi^-}$ in $^{16}_{\Xi^-}$C
due to a high momentum transfer $q_{\Xi^-}\simeq$ 400 MeV/c.
The $\Xi^-$ admixture probabilities of these states are on the order of 5--9\%.
However, populations of the $0^+$ ground state with 
${^{14}{\rm C}(0^+)}\otimes s_{\Lambda}^2$ and the $2^+$ excited state with 
${^{14}{\rm C}(2^+)}\otimes s_{\Lambda}^2$ are very small. 
The sensitivity of the spectrum on the $\Xi N\text{--}\Lambda\Lambda$ 
coupling strength
enables us to extract the nature of $\Xi N\text{--}\Lambda\Lambda$ dynamics
in nuclei, and the nuclear ($K^-$,~$K^+$) reaction can extend our knowledge 
of the $S=-2$ world.

\end{abstract}
\maketitle


\section{Introduction}

It is important to understand properties of $\Xi$ hypernuclei
whose states are regarded as ``doorways'' to access 
multi-strangeness systems as well as 
a two-body $\Xi N$--$\Lambda\Lambda$ system,
and it is a significant step to extend study of strange 
nuclear matter in hadron physics and astrophysics \cite{NPA804}.
Because the $\Xi$ hyperon in nuclei has to undergo 
a strong $\Xi N \to \Lambda\Lambda$ decay, 
widths of $\Xi$ hypernuclear states give us 
a clue to a mechanism of $\Xi$ absorption processes 
in nuclei.
A pioneer study of $\Xi$ hypernuclei by Dover and Gal \cite{Dover83}
has found that a $\Xi$-nucleus 
potential has a well depth of $24 \pm 4$ MeV in the real part 
on the analysis of old emulsion data.
However, 
our knowledge of these $\Xi$-nucleus systems is very limited 
due to the lack of the experimental data \cite{NagaeE05}.
Indeed, the missing-mass spectra of a double-charge 
exchange (DCX) reaction ($K^-$,~$K^+$) on a $^{12}$C target
have suggested the $\Xi$ well depth of 
14--16 MeV \cite{Fukuda98,Khaustov00}.
Several authors \cite{Tadokoro95} have used the unsettled 
$\Xi$-nucleus (optical) potentials such as 
$V_\Xi=$($-$24)--($-$14) MeV and $W_\Xi=$($-$6)--($-$3) MeV in the 
Woods-Saxon potential to demonstrate the $\Xi^-$ production 
spectra in the nuclear ($K^-$, $K^+$) reactions.
There remains a full uncertainty about the nature of doubly 
strange ($S=-2$) dynamics caused by the $\Xi N$ and 
$\Xi N$--$\Lambda\Lambda$ interaction in nuclei 
at the present stage.
More experimental information is earnestly desired.

\begin{figure}[tb]
  \begin{center}
  \includegraphics[width=0.60\linewidth]{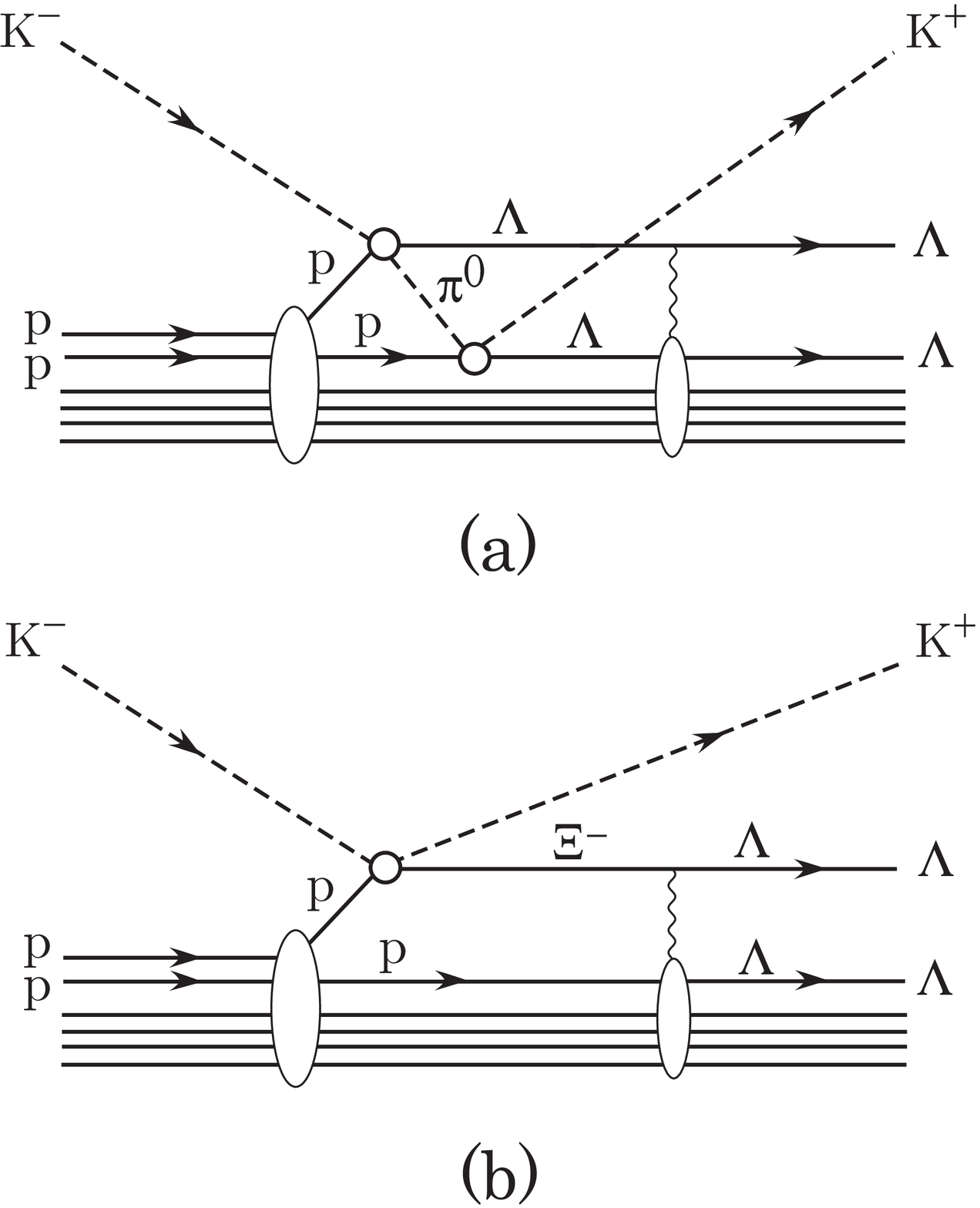}
  \caption{\label{fig:1}
  Diagrams for DCX nuclear ($K^-$,~$K^+$) reactions:  
  (a) a two-step mechanism, $K^-p \to \pi^0\Lambda$ followed by 
  $\pi^0p \to K^+\Lambda$, and (b) a one-step mechanism, 
  $K^-p \to K^+ \Xi^-$ via $\Xi^-$ doorways caused by 
  the $\Xi^-p\text{--}\Lambda \Lambda$ coupling.
  }
  \end{center}
\end{figure}

The $(K^-$,~$K^+$) reaction 
is one of the most promising ways of studying doubly 
strange systems such as $\Xi^-$ hypernuclei 
for the forthcoming 
J-PARC experiments \cite{NagaeE05}. 
One expects that these experiments will confirm the existence of 
$\Xi$ hypernuclei and establish 
properties of the $\Xi$-nucleus potential, 
e.g., binding energies and widths. 
This reaction can also populate a
$\Lambda\Lambda$ hypernucleus through a conventional 
DCX two-step mechanism as $K^-p \to \pi^0\Lambda$ followed by 
$\pi^0 p \to K^+\Lambda$ \cite{Dover80,Baltz83,Iijima92}, 
as shown in Fig.~\ref{fig:1}(a). 
Such an inclusive $K^-$ spectrum in the $\Lambda\Lambda$ bound 
region is rather clean with much less background experimentally.
Early theoretical predictions for two-step 
$^{16}{\rm O}$($K^-$, $K^+$) reactions at the incident momentum 
$p_{K^-}=$ 1.1 GeV/c and scattering angle 
$\theta_{\rm lab}=0^\circ$ \cite{Dover80,Baltz83} have 
indicated small cross sections for the $\Lambda\Lambda$
states, for example, 
$\sim$0.1 nb/sr for the $0^+(s^2_\Lambda)$ ground state and 
$\sim$2 nb/sr for the $2^+(s^2_\Lambda)$ excited state 
in $^{~16}_{\Lambda\Lambda}{\rm C}$
when we took 0.61 mb/sr and 0.32 mb/sr as 
the laboratory cross sections at $0^{\circ}$ for
$K^-p \to \pi^0\Lambda$ and $\pi^0 p \to K^+\Lambda$, 
respectively.

It should be noticed that another exotic production 
of $\Lambda\Lambda$ hypernuclei in the ($K^-$,~$K^+$) reactions
is a one-step mechanism, 
$K^-p \to K^+\Xi^-$ via $\Xi^-$ doorways caused 
by a $\Xi^-p \to \Lambda\Lambda$ transition,
as shown in Fig.~\ref{fig:1}(b).
The $\Xi N$--$\Lambda\Lambda$ coupling induces the
$\Xi^-$ admixture and the $\Lambda\Lambda$ energy shift  
$\Delta B_{\Lambda\Lambda}\equiv
B_{\Lambda\Lambda}({^{~A}_{\Lambda\Lambda}{\rm Z}})
-2B_{\Lambda}({^{A-1}_{~\Lambda}{\rm Z}})$
in the $\Lambda\Lambda$-nuclear 
states \cite{Myint94,Myint03,Carr97,Yamada00,Lanskoy04},
and its coupling strength is also related to widths of 
$\Xi$-hypernuclear 
states \cite{Ikeda94,Dover94}.
For a viewpoint of $S=-2$ studies, it is very important to extract 
quantitative information concerning 
the $\Xi N$--$\Lambda\Lambda$ coupling 
from spectroscopy of the $\Xi$ and 
$\Lambda\Lambda$ hypernuclei \cite{Takahashi01,Hiyama02}.

\Figuretable{FIG. 1}

In this Letter, we study theoretically production of 
a doubly strange hypernucleus 
in the DCX ($K^-$,~$K^+$) reaction 
on an $^{16}$O target at $p_{K^-}=$ 1.8 GeV/c and 
$\theta_{\rm lab}=$ 0$^\circ$
within a distorted-wave impulse approximation (DWIA).
Thus we focus on the $\Lambda\Lambda$--$\Xi$ 
spectrum for $^{~16}_{\Lambda\Lambda}$C and 
$^{~16}_{~\Xi^-}$C in the $\Xi^-$ bound region
considering the one-step mechanism, 
$K^- p \to K^+ \Xi^-$ via $\Xi^-$ doorways caused by 
the $\Xi N$--$\Lambda \Lambda$ coupling 
in the nuclear ($K^-$,~$K^+$) reaction,
rather than the two-step mechanism as 
$K^- p \to \pi^0 \Lambda$ followed by 
$\pi^0 p \to K^+ \Lambda$ \cite{Dover80,Baltz83}.
These different mechanisms are well separated kinematically. 
The forward cross section for the $K^- p \to K^+ \Xi^-$ elementary 
process is at its maximum at $p_{K^-}=$ 1.8--1.9 GeV/c, 
whereas the $K^- p \to \pi^0 \Lambda$ reaction at $p_{K^-}=$ 1.1 GeV/c 
leads to the maximal cross section for the 
$\pi^0 p \to K^+ \Lambda$ process. 
The present study is the first attempt to 
evaluate a production spectrum of the $\Lambda\Lambda$--$\Xi$ 
hypernucleus via the $\Xi N$--$\Lambda\Lambda$ coupling
from the inclusive ($K^-$,~$K^+$) reaction, 
and to extract the $\Xi^-$ admixture probability in the $\Lambda\Lambda$ 
hypernucleus from the spectrum.
We also discuss a contribution of the two-step
processes in the ($K^-$,~$K^+$) reactions within 
the eikonal approximation.

\section{Calculations}

Let us consider the DCX ($K^-$,~$K^+$) reaction on 
the $^{16}$O target at 1.8 GeV/c within a DWIA
and examine the production cross sections and wave functions 
of the doubly strange hypernucleus.
To fully describe the one-step process via $\Xi^-$ doorways, 
as shown in Fig.~\ref{fig:1}(b), 
we perform nuclear $\Lambda\Lambda$--$\Xi$ coupled-channel 
calculations \cite{Myint03,Lanskoy04}, 
which are assumed to effectively represent the coupling nature 
in omitting other $\Lambda\Sigma$ and 
$\Sigma\Sigma$ channels for simplicity. 
Here we employ a multichannel coupled wave function of 
the $\Lambda\Lambda$--$\Xi$ 
nuclear state for a total spin $J_B$ within 
a weak coupling basis.
It is written as 
\begin{eqnarray}
|\varPsi_{J_B}(^{~~~16}_{\Lambda\Lambda\mbox{-}\Xi}{\rm C}) \rangle
&=& \sum_{JJ''j_1j_2}
\big[[\Phi_{J}({^{14}{\rm C}}),
\varphi^{(\Lambda)}_{j_1}({\bm r}_{\Lambda_1})]_{J''},
\varphi^{(\Lambda)}_{j_2}({\bm r}_{\Lambda_2})\big]_{J_B}
\nonumber \\
&+& \sum_{JJ'j_pj_3}
\big[\Phi_{J'}({^{15}{\rm N}}),\varphi^{(\Xi^-)}_{j_3}({\bm r}_\Xi)\big]_{J_B} 
\label{eqn:e1}
\end{eqnarray}
with 
$\Phi_{J'}({^{15}{\rm N}})= 
{\cal A}[\Phi_{J}({^{14}{\rm C}}),\varphi^{(p)}_{j_p}({\bm r}_p)]_{J'}$, 
where 
${\bm r}_{\Lambda_1}$ (${\bm r}_p$) denotes the relative coordinate between 
the ${^{14}{\rm C}}$ core-nucleus and the $\Lambda$ (proton), 
and ${\bm r}_{\Lambda_2}$ (${\bm r}_\Xi$) denotes the relative coordinate
between the center of mass of the $^{14}{\rm C}$--$\Lambda$ ($^{15}{\rm N}$) 
subsystem and the $\Lambda$ ($\Xi^-$). 
Thus $\varphi^{(\Lambda)}_{j_{1,2}}$, $\varphi^{(\Xi^-)}_{j_{3}}$ 
and $\varphi^{(p)}_{j_{p}}$ describe the relative wave functions 
of shell model states 
(that occupy $j_{1,2}$, $j_3$ and $j_p$ orbits)
for the $\Lambda$, $\Xi^-$ and proton, respectively; 
$\Phi_J({^{14}{\rm C}})$ is a wave function of the ${^{14}{\rm C}}$ 
core-nucleus state,
and ${\cal A}$ is the anti-symmetrized operator for nucleons. 
The energy difference between 
$^{15}{\rm N}+\Xi^-$ and ${^{14}{\rm C}}+\Lambda+\Lambda$ channels
is 
$\Delta M= M({^{15}{\rm N}})+m_{\Xi^-}
-M({^{14}{\rm C}})-2m_{\Lambda}=$ 18.4 MeV, 
where $M({^{15}{\rm N}})$, $M({^{14}{\rm C}})$, 
$m_{\Xi^-}$ and $m_{\Lambda}$ are
masses of the ${^{15}{\rm N}}$ nucleus, 
the ${^{14}{\rm C}}$ nucleus, the $\Xi^-$ and $\Lambda$ hyperons,
respectively.
We take the ${^{15}{\rm N}}$ core-nucleus states 
with $J^\pi={1/2}^-$(g.s.) and ${3/2}^-$(6.32 MeV),  
and the ${^{14}{\rm C}}$ core-nucleus states 
with $J^\pi={0}^+$(g.s.) and ${2}^+$(7.01 MeV) that are given 
in $(0p^{-1}_{1/2}0p^{-1}_{1/2})_{0^+}$, $(0p^{-1}_{3/2}0p^{-1}_{1/2})_{2^+}$
and $(0p^{-1}_{3/2}0p^{-1}_{3/2})_{0^+,2^+}$
configurations on ${^{16}{\rm O(g.s.)}}$ \cite{Dover80,Baltz83}.
Because we assume only natural-parity $\pi=(-1)^{J_B}$ states 
via $\Xi^-$ doorways that are selectively formed by non-spin-flip processes 
in the forward $K^-p \to K^+\Xi^-$ reaction, 
we consider a spin $S=$ 0, $\Lambda\Lambda$ pair in the hypernucleus.
If the $\Lambda\Lambda$ component is dominant in a bound state, 
we can identify it as a state of the $\Lambda\Lambda$ hypernucleus
$^{~16}_{\Lambda\Lambda}{\rm C}$, in which the 
$\Xi^-$ admixture probability can be estimated by
\begin{equation}
P_{\Xi^-}= \sum_{j_pj_3}\langle {\varphi}^{(p)}_{j_p}
{\varphi}^{(\Xi^-)}_{j_3}|\varphi^{(p)}_{j_p} \varphi^{(\Xi^-)}_{j_3} \rangle,
\end{equation}
under the normalization of 
$\sum_{j_1j_2}\langle {\varphi}^{(\Lambda)}_{j_1} {\varphi}^{(\Lambda)}_{j_2}| 
          \varphi^{(\Lambda)}_{j_1} \varphi^{(\Lambda)}_{j_2} \rangle
+\sum_{j_pj_3}\langle {\varphi}^{(p)}_{j_p} {\varphi}^{(\Xi^-)}_{j_3}| 
 \varphi^{(p)}_{j_p} \varphi^{(\Xi^-)}_{j_3} \rangle$ =1.

After we set up the 
${{^{15}_{\Lambda}}{\rm C}}$ and $^{15}$N
configurations in our model space with Eq.~(\ref{eqn:e1}), 
we calculate the wave functions of 
$\varphi^{(\Lambda)}_{j_2}({\bm r}_{\Lambda_2})$ and 
$\varphi^{(\Xi^-)}_{j_3}({\bm r}_\Xi)$ taking into account 
their channel coupling.  
Thus, the complete Green's function ${\bm G}(\omega)$ \cite{Morimatsu94} 
describes all information concerning 
$({^{15}_{\Lambda}}{\rm C} \otimes \Lambda)
+({^{15}}{\rm N} \otimes \Xi^-)$ coupled-channel dynamics,
as a function of the energy transfer $\omega$. 
It is numerically obtained as a solution of the 
$N$-channels radial coupled equations
with a hyperon-nucleus potential ${\bm U}$ \cite{Harada98,Harada09}, 
which is written in an abbreviated notation as
\begin{equation}
{\bm G}(\omega)={\bm G}^{(0)}(\omega)
+{\bm G}^{(0)}(\omega){\bm U}{\bm G}(\omega)
\label{eqn:e2}
\end{equation}
with 
\begin{equation}
{\bm G}(\omega)=\left(
\begin{array}{cc}
    {G}_{\Lambda}(\omega) & {G}_{X}(\omega)  \\
    {G}_{X}(\omega) & {G}_{\Xi}(\omega)   
\end{array}
\right), \quad
{\bm U}=\left(
\begin{array}{cc}
    {U}_{\Lambda} & {U}_{X} \\
    {U}_{X} & {U}_{\Xi}   
\end{array}
\right), 
\label{eqn:e3}
\end{equation}
where ${\bm G}^{(0)}(\omega)$ is a free Green's function.
In our calculations, for example, we deal with $N=28$ for 
the $J_B^\pi=$ $1^-$ state.
The nuclear optical potentials 
$U_{Y}$ ($Y=$ $\Xi$ or $\Lambda$) can be written as  
\begin{equation}
U_{Y}(r)= V_{Y}f(r,R,a)+iW_{Y}f(r,R',a')
+iW^{(D)}_{Y}g(r,R',a'),
\label{eqn:e4}
\end{equation}
where $f$ is the Woods-Saxon (WS) form, 
$f(r,R,a)=[1 + \exp{((r-R)/a)}]^{-1}$, 
and $g$ is the derivative of the WS form, 
$g(r,R',a')=-4a'(d/dr)f(r,R',a')$.
The spin-orbit potentials are neglected.
In $^{15}{\rm N}$--$\Xi^-$ channels, we assume 
the strength parameter of $V_\Xi=$ $-24$ or $-14$ MeV 
with $a=$ 0.6 fm and $R=1.10(A-1)^{1/3}=$ 2.71 fm 
in $U_{\Xi}(r)$ \cite{Dover83,Tadokoro95,Khaustov00},
taking into account the Coulomb potential with the 
nuclear finite size $R_C=1.25A^{1/3}=$ 3.15 fm \cite{Bohr69}.
The spreading imaginary potential in Eq.~(\ref{eqn:e4}), 
${\rm Im}\,U_Y$, expresses complicated excited states 
via $\Xi^-N \to \Lambda\Lambda$ conversion processes in 
${^{~16}_{\Xi^-}}{\rm C}$ or ${^{~16}_{\Lambda\Lambda}}{\rm C}$
above the $^{15}_{\Lambda\Lambda}{\rm C} + n$ threshold at 8.2 MeV, 
as a function of the excitation energy 
$E_{\rm ex}$ measured from an energy of 
the $^{~16}_{\Lambda\Lambda}{\rm C}$ ground state, 
as often used in nuclear optical models.
Since we have no criterion for a choice of $W_{\Xi}$ or $W^{(D)}_{\Xi}$ 
in the limited experimental data, 
we adjust appropriately the strength parameter of $W_{\Xi}$ in the WS-type 
to give widths of $\Xi^-$ quasibound states 
in recent calculations \cite{Tadokoro95,Khaustov00,Hiyama08}. 
In $^{14}{\rm C}$--$\Lambda\Lambda$ channels, we should use
a $^{15}_\Lambda{\rm C}$--$\Lambda$ potential, which can be 
constructed in folded potential models \cite{Satchler83}:
\begin{equation}
U_{\Lambda}(r)=\int
\rho_{J''}({\bm r}_{\Lambda})
\big[
U_{{\rm C}\Lambda}(|{\bm r}+ \lambda_\Lambda{\bm r}_{\Lambda}|)
+V_{\Lambda\Lambda}(|{\bm r}- \nu_\Lambda{\bm r}_{\Lambda}|)
\big]d{\bm r}_{\Lambda},
\label{eqn:e4b}
\end{equation}
where $\rho_{J''}({\bm r}_{\Lambda})
=\sum_{j_1m_1}\langle JMj_1m_1|J''M''\rangle^2
|\varphi^{(\Lambda)}_{j_1}({\bm r}_{\Lambda})|^2$ and 
$\lambda_\Lambda=1-\nu_\Lambda=
m_\Lambda/(M({^{14}{\rm C}})+m_\Lambda)$. 
$U_{{\rm C}\Lambda}$ and $V_{\Lambda\Lambda}$ 
denote an optical potential for $^{14}{\rm C}$--$\Lambda$ 
as given in Eq.~(\ref{eqn:e4})
and a $\Lambda\Lambda$ residual interaction, respectively. 
Here we neglected $V_{\Lambda\Lambda}$ for simplicity.
The real part of $U_{{\rm C}\Lambda}$
leads to $B_{\Lambda}=$ 12.2 MeV for the $(0s)_\Lambda$ state 
and $B_{\Lambda}=$ 1.6 MeV for the $(0p)_\Lambda$ state 
in $^{15}_{\Lambda}{\rm C}$ \cite{Millener88}, 
and its imaginary part exhibits 
a flux loss of the wave functions 
through the core excitations of $^{14}{\rm C}^*$.
We assume $W_{\Lambda}\simeq {1 \over 4}W_N$ and 
$W^{(D)}_{\Lambda}\simeq {1 \over 4}W^{(D)}_N$ 
where parameters of $W_N$ and $W^{(D)}_N$ for nucleon 
were obtained in Ref.~\cite{Watson69} 
because the well depth of the imaginary potential for $\Lambda$ 
is by a factor of 4 weaker than that for nucleon 
in $g$-matrix calculations \cite{Yamamoto88}. 
 
The $\Lambda\Lambda$--$\Xi$ coupling potential 
$U_{X}$ in off-diagonal parts of ${\bm U}$ 
is the most interesting object in this calculation
\cite{Myint94,Carr97,Yamada00,Myint03,Lanskoy04,Ikeda94,Dover94}.
It can be obtained by a two-body $\Xi N$--$\Lambda\Lambda$ potential 
$v_{\Xi N,\Lambda\Lambda}({\bm r}',{\bm r})$ with the $^1S_0$, isospin 
$T=$ 0 state. 
Here we use a zero-range interaction 
$v_{\Xi N,\Lambda\Lambda}({\bm r}',{\bm r})= 
v^0_{\Xi N,\Lambda\Lambda}\delta_{S,0}\delta({\bm r}'-{\bm r})$ 
in a real potential for simplicity, 
where $v^0_{\Xi N,\Lambda\Lambda}$ is the strength 
parameter that should be connected with volume integral 
$\int v_{\Xi N,\Lambda\Lambda}({\bm r}) d{\bm r}=v^0_{\Xi N,\Lambda\Lambda}$
\cite{Dover94,Myint03,Lanskoy04}.
Thus the matrix elements can be easily estimated by use of 
Racah algebra \cite{Glendenning83}: 
\begin{eqnarray}
U_{X}(r) 
&=& \Big\langle \big[ \Phi_{J'}({^{15}{\rm N}})
\otimes {\cal Y}^{(\Xi^-)}_{j'\ell' s'}(\hat{\bm r}) \big]_{J_B}|
\sum_i v_{\Xi N,\Lambda\Lambda}(\nu_i{\bm r}'_i,{\bm r})  \nonumber\\
&&\times 
\big| \big[ [\Phi_J({^{14}{\rm C}}),
\varphi^{(\Lambda)}_{j_1}]_{J''}
\otimes {\cal Y}^{(\Lambda)}_{j\ell s}(\hat{\bm r}) \big]_{J_B} 
\Big\rangle \nonumber\\
&=& 
\sum_{LSK} \sqrt{1/2}\,v^0_{\Xi N,\Lambda\Lambda}
\delta_{S,0}C^{J_B}_{LSK}(J'J'')
{\cal F}^{J'J''}_{LSK}(r),
\label{eqn:e5}
\end{eqnarray}
where ${\cal Y}_{j \ell  s}=[Y_\ell \otimes X_{1 \over 2}]_j$
is a spin-orbit function and 
$C^{J_B}_{LSK}(J'J'')$ is a purely geometrical factor \cite{Glendenning83};
${\cal F}^{J'J''}_{LSK}(r)$ is the nuclear form 
factor including a recoupling coefficient of 
$U(Jj_1{J''}K;{J'}j_p)$ \cite{Dover94}, a parentage coefficient 
for proton removal 
from ${^{15}{\rm N}}(1/2^-,3/2^-)$ \cite{Cohen67} and the 
center-of-mass correction of a factor $\sqrt{A/(A-1)}$ \cite{Dieperink74}. 
The factor $\sqrt{1/2}$ comes from the procedure handling 
a transition between $p \Xi^-$ and $\Lambda\Lambda$ states in the nucleus. 

The inclusive $K^+$ double-differential laboratory cross section 
of the $\Lambda\Lambda$--$\Xi$ production in the nuclear 
($K^-$,~$K^+$) reaction  
can be written within the DWIA \cite{Hufner74,Auerbach83} 
using the Green's function method \cite{Morimatsu94}. 
In the one-step mechanism, $K^- p \to K^+ \Xi^-$ via $\Xi^-$ 
doorways, it is given \cite{Harada09} as
\begin{eqnarray}
&&\left({{d^2\sigma} 
\over {d\Omega_{K} dE_{K}} }\right)_{\rm lab} \nonumber\\
&&= \beta {1 \over {[J_A]}} \sum_{M_z}
\sum_{\alpha' \alpha}(-{1 \over \pi}) 
{\rm Im} \Big[ \int \! d{\bm r}'\!d{\bm r}\,
F_{\Xi}^{\alpha' \dagger}({\bm r}')   \nonumber\\
&& \quad \times \ {G}_{\Xi}^{\alpha'\alpha}(\omega,{\bm r}',{\bm r})
F_{\Xi}^{\alpha \,}({\bm r})
\Big]
\label{eqn:e1a}
\end{eqnarray}
for the target with a spin $J_A$ and its $z$-component $M_z$,
where $[J_A]=2J_A+1$, and a kinematical factor $\beta$ \cite{Koike08}
that expresses the translation 
from the two-body $K^-$--$p$ laboratory system to the $K^-$--$^{16}$O
laboratory system \cite{Dover83}. 
The production amplitude $F_{\Xi}^{\alpha}$ is 
\begin{eqnarray}
  F_{\Xi}^{\alpha}({\bm r}) &=& {\overline{f}}_{K^-p \to K^+\Xi^-} 
  \chi_{{\bm p}_{K^+}}^{(-) \ast}({M_C \over M_B}{\bm r})
  \chi_{{\bm p}_{K^-}}^{(+)}({M_C \over M_A}{\bm r}) \nonumber\\
 && \times 
  \langle \alpha \, | \hat{\mit\psi}_p({\bm r})| \, \varPsi_{J_AM_z} \rangle,
\label{eqn:e1b}
\end{eqnarray}
where 
$\overline{f}_{K^-p \to K^+\Xi^-}$ is a  
Fermi-averaged amplitude for the $K^-p \to K^+\Xi^-$ reaction 
in nuclear medium \cite{Dover83}, and 
$\chi_{{\bm p}_{K^+}}^{(-)}$ and $\chi_{{\bm p}_{K^-}}^{(+)}$ 
are the distorted waves for outgoing $K^+$ and 
incoming $K^-$ mesons, respectively;
the factors of ${M_C/M_B}$ and ${M_C/M_A}$ take into account 
the recoil effects, where 
$M_A$, $M_B$ and $M_C$ are masses of the target, the final state
and the core-nucleus, respectively. 
$\langle \alpha \, | \hat{\mit\psi}_p  | \, \varPsi_{J_AM_z} \rangle$ 
is a hole-state wave function for a struck proton in the target, 
where $\alpha$ denotes the complete set of eigenstates for the system.
It should be recognized that the $\Lambda\Lambda$--$\Xi$ 
coupled-channel Green's function with the spreading potential 
provides an advantage of estimating contributions from sources both 
as $\Lambda\Lambda$ components in $\Xi^-$-nucleus eigenstates \cite{Dover94} 
and as $\Xi^-p\to\Lambda\Lambda$ quasi-scattering processes 
in the nucleus \cite{Ikeda94}.

Because the momentum transfer is very high in 
the nuclear ($K^-$,~$K^+$) reaction 
at 1.8 GeV/c, i.e., $q_{\Xi^-} \simeq$ 360--430 MeV/c, 
the distorted waves for outgoing $K^+$ and 
incoming $K^-$ in Eq.~(\ref{eqn:e1b}) are calculated 
with the help of the eikonal approximation \cite{Hufner74,Harada04}.
As the distortion parameters, we use total cross sections 
of $\sigma_{K^-N}$= 28.9 mb for $K^- N$ scattering and 
$\sigma_{K^+N}$= 19.4 mb for $K^+ N$ scattering \cite{Tadokoro95}, 
and $\alpha_{K^-N} = \alpha_{K^+N} =$ 0.
We take 35 $\mu$b/sr as the laboratory cross section of 
$d\sigma/d\Omega = \bar{\alpha}|{\overline{f}}_{K^-p \to K^+\Xi^-}|^2$
including the kinematical factor $\bar{\alpha}$ \cite{Iijima92,Khaustov00}.
For the target nucleus $^{16}$O with $J^\pi_A=0^+$, 
we assume the wave functions for the proton hole-states in the relative 
coordinate, which are calculated 
with central (WS-type) and spin-orbit potentials \cite{Bohr69}, 
by fitting to the charge rms radius of 2.72 fm \cite{Vries87}.
For the energies (widths) for proton-hole states, 
we input 12.1 (0.0), 18.4 (2.5) and 36 (10) MeV for 
${0p}_{1/2}^{-1}$, ${0p}_{3/2}^{-1}$ 
and $0s_{1/2}^{-1}$ states, 
respectively.

Three parameters, $V_\Xi$, $W_\Xi$ and 
$v^0_{\Xi N,\Lambda\Lambda}$,  are very important for 
calculating the inclusive 
spectra with the one-step mechanism.
These parameters are strongly connected each other 
for the shape of the spectrum and its magnitude, 
as well as for the $\Xi^-$ binding energies and widths 
of the $\Xi^-$ states. 
Several authors \cite{Myint94,Dover94,Lanskoy04} investigated 
the effects of the $\Xi N$--$\Lambda\Lambda$ coupling in light nuclei 
evaluating the volume integrals 
for $k_F$-dependent $\Xi N$--$\Lambda\Lambda$ effective interactions 
based on Nijmegen potentials \cite{Yamamoto08}, in which 
these values are strongly model dependent; for example,  
250.9, 370.2, 501.5, 582.1 and 873.9 MeV$\cdot$fm$^3$ for NHC-D, NSC97e, 
NSC04a, NHC-F and NSC04d potentials ($k_F=$ 1.0 fm$^{-1}$), 
respectively \cite{Lanskoy04,Yamamoto08}.
The $\Xi^-p \to \Lambda\Lambda$ conversion cross section
of $(v \sigma)_{\Xi^-p \to \Lambda\Lambda}\simeq$ 7.9 mb 
also yields to be about 544 MeV$\cdot$fm$^3$ \cite{Dover94}. 
To see the dependence of the spectrum on the 
$\Xi N$--${\Lambda\Lambda}$ coupling strength, here, 
we choose typical values of $v^0_{\Xi N,\Lambda\Lambda}=$ 
250 and 500 ${\rm MeV}$, which approximate the volume 
integrals of NHC-D and NSC04a, respectively. 
We take the spreading potential of ${\rm Im}\,U_{\Xi}$ 
to be $W_{\Xi}\simeq$ $-3$ MeV at the $^{15}{\rm N}$ + $\Xi^-$ threshold 
\cite{Tadokoro95,Khaustov00,Lanskoy04,Hiyama02}.
It should be noticed that this spreading potential
expresses nuclear core breakup processes caused by the 
$\Xi^- p \to \Lambda\Lambda$ conversion in the $^{15}$N nucleus,
and its effect cannot be involved in $U_X$.

\section{Results and discussion}

Now let us discuss the inclusive spectrum in the 
$^{16}{\rm O}$($K^-$,~$K^+$) reaction at 1.8 GeV/c (0$^\circ$)
in order to examine the dependence of the spectrum on 
the parameters of $V_\Xi$ and $v^0_{\Xi N,\Lambda\Lambda}$. 
We consider contributions of the $\Lambda\Lambda$--$\Xi$ 
nuclear bound and resonance states to the $\Xi^-p\to \Lambda\Lambda$
conversion processes in the $\Xi^-$ bound region.

\begin{figure}[tb]
  \begin{center}
 \includegraphics[width=0.75\linewidth]{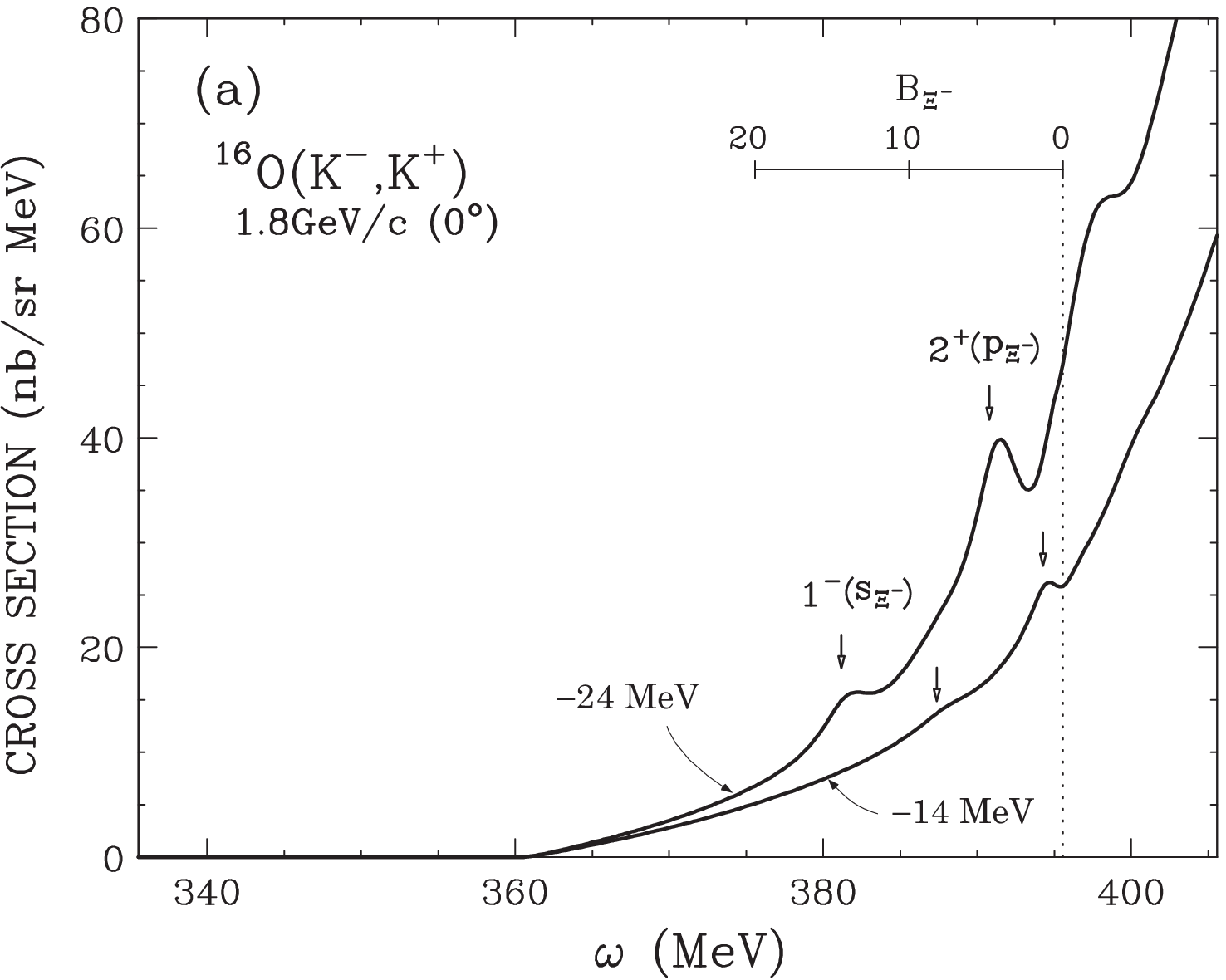}
 \includegraphics[width=0.75\linewidth]{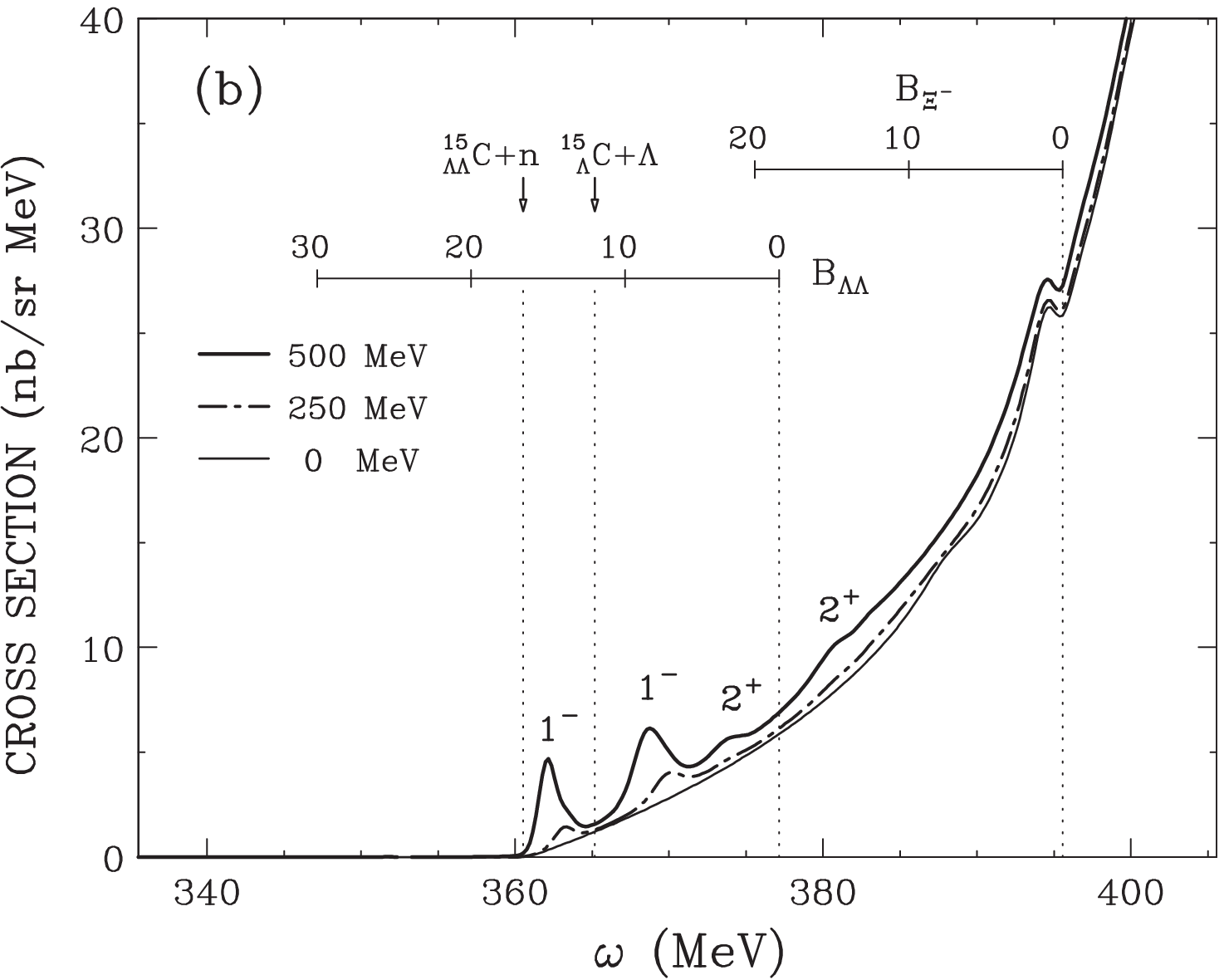}
  \caption{\label{fig:2}
  Calculated inclusive $\Lambda\Lambda\text{--}\Xi$ 
  spectra by the one-step mechanism in the $^{16}$O($K^-$,~$K^+$) 
  reaction at 1.8 GeV/c (0$^\circ$), with a detector resolution 
  of 1.5 MeV FWHM; 
  (a) $V_{\Xi}=$ $-$24 or $-14$ MeV without the $\Lambda\Lambda\text{--}\Xi$ 
  coupling potential.
  The $\Xi$ conversion decay occurs above the $^{~15}_{\Lambda\Lambda}{\rm C}+n$ 
  threshold at $\omega=$ 361 MeV; 
  (b) $V_{\Xi}=$ $-14$ MeV with the $\Lambda\Lambda\text{--}\Xi$ 
  coupling potential obtained by
  $v^0_{\Xi N,\Lambda\Lambda}=$ 0, 250 and 500 MeV. 
  }
  \end{center}
\end{figure}

In Fig.~\ref{fig:2}(a), we show the calculated spectra in 
the $\Xi^-$ bound region without the $\Lambda\Lambda $--$\Xi$ coupling 
potential when we use $V_{\Xi}=$ $-24$ MeV or $-14$ MeV with 
the Coulomb potential.
The calculated spectra are in agreement with the spectra 
obtained by previous works~\cite{Tadokoro95}. 
In the case of $V_{\Xi}=$ $-24$ MeV, we find that 
a broad peak of the 
$[{^{15}{\rm N}{\rm (1/2^-)}}\otimes s_{\Xi^-}]_{1^-}$ quasibound state 
in $^{16}_{\Xi^-}{\rm C}$ 
is located at $B_{\Xi^-}=$ 13.4 MeV with a sizable width of 
$\varGamma=$ 3.5 MeV, and a clear peak of the 
$[{^{15}{\rm N}{\rm (1/2^-)}}\otimes p_{\Xi^-}]_{2^+}$ quasibound state 
at $B_{\Xi^-}=$ 3.7 MeV with $\varGamma=$ 3.1 MeV. 
Integrated cross sections indicate 
$d\sigma(0^\circ)/d\Omega \simeq$ 28 nb/sr for the $1^-$ state and 
77 nb/sr for the $2^+$ state in $^{16}_{\Xi^-}{\rm C}$.
In the case of $V_{\Xi}=$ $-14$ MeV, which is favored in recent calculations
\cite{Myint03,Lanskoy04,Tadokoro95,Hiyama02}, we have 
the $[{^{15}{\rm N}{\rm (1/2^-)}}\otimes s_{\Xi^-}]_{1^-}$  state 
at $B_{\Xi^-}=$ 6.8 MeV with $\varGamma=$ 3.8 MeV
and the $[{^{15}{\rm N}{\rm (1/2^-)}}\otimes p_{\Xi^-}]_{2^+}$ 
at $B_{\Xi^-}=$ 0.5 MeV with $\varGamma=$ 1.1 MeV.
The integrated cross sections indicate 
$d\sigma(0^\circ)/d\Omega \simeq$ 6 nb/sr 
for the $1^-$ state and 9 nb/sr for the $2^+$ state.
Note that the $\Xi^-p \to \Lambda\Lambda$ conversion processes 
that can be described by the absorption potential ${\rm Im}\,U_{\Xi}$, 
must appear above the $^{~15}_{\Lambda\Lambda}{\rm C}+n$ decay threshold 
at $\omega=$ 360.4 MeV (which corresponds to 
$B_{\Lambda \Lambda}=$ 16.7 MeV). 
We confirm that no clear signal of the $\Xi^-$ bound state 
is measured 
if $V_\Xi$ is sallow such as $-V_\Xi \leq$ 14 MeV 
and/or $W_\Xi$ is sizably absorptive 
($-W_\Xi \geq$ 3 MeV at the 
${^{15}{\rm N}}+\Xi^-$ threshold) in $U_\Xi$. 
Nevertheless, the production of these $\Xi^-$ states 
as well as $\Xi^-$ states coupled to a ${^{15}{\rm N}(3/2^-)}$ nucleus 
is essential in this model 
because these states act as doorways when we consider the 
$\Lambda\Lambda$ states formed in the one-step mechanism. 
We also expect to extract properties 
of the $\Xi$-nucleus potential such as $V_\Xi$ and $W_\Xi$ 
from the $\Xi^-$ continuum spectra in the ($K^-$, $K^+$) 
reactions on nuclear targets, as already discussed for studies 
of the $\Sigma^-$-nucleus potential in nuclear 
($\pi^-$,~$K^+$) reactions \cite{Friedman07,Harada05}.

\Figuretable{FIG. 2}

On the other hand, the $\Lambda\Lambda $--$\Xi$ coupling 
plays an important role in making a production of 
the $\Lambda\Lambda$ states via $\Xi^-$ doorways 
below the ${^{15}{\rm N}+\Xi^-}$ threshold.
The positions of their peaks 
must be slightly shifted downward by the energy shifts 
$\Delta B_{\Lambda\Lambda}$ 
due to the coupling potential in Eq.~(\ref{eqn:e5}).
When $v^0_{\Xi N,\Lambda\Lambda}=$ 500 MeV (250 MeV),
we obtain $\Delta B_{\Lambda\Lambda}=$ 1.17 MeV (0.15 MeV) and 
the $\Xi^-$ admixture probability $P_{\Xi^-}=$ 5.24\% (0.87\%) in 
the $[{^{14}{\rm C}(0^+)}\otimes s_\Lambda p_\Lambda]_{1^-}$ 
excited state and 
$\Delta B_{\Lambda\Lambda}=$ 0.38 MeV (0.09 MeV) and 
$P_{\Xi^-}=$ 0.58\% (0.14\%) in 
the $[{^{14}{\rm C}(0^+)}\otimes s^2_\Lambda]_{0^+}$ 
ground state.
The value of $P_{\Xi^-}$ in the $1^-$ state is by a factor of 
6--9 as large as that in the $0^+$ state.
These values are strongly connected with the magnitude of the peak for 
the $\Lambda\Lambda$ state in the spectrum.

\begin{figure}[t]
  \begin{center}
 \includegraphics[width=0.8\linewidth]{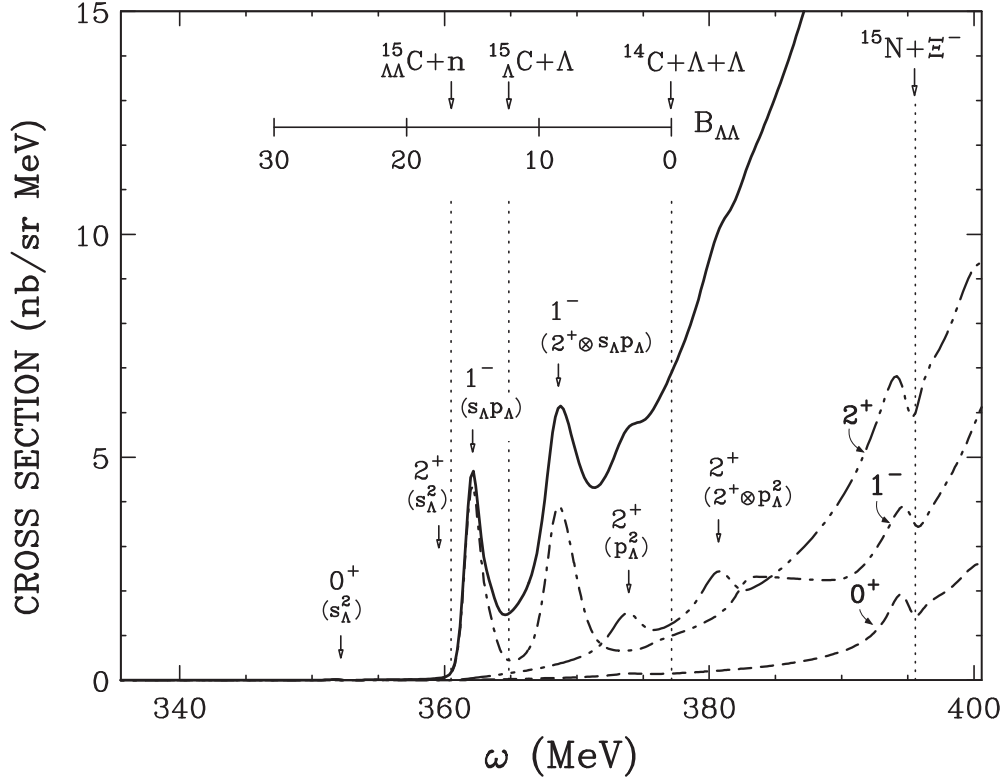}
  \caption{\label{fig:3}
  Partial-wave decomposition of the calculated inclusive spectrum 
  by the one-step mechanism 
  near the $^{14}$C+$\Lambda$+$\Lambda$ threshold 
  in the $^{16}$O($K^-$,~$K^+$) reaction at 1.8 GeV/c (0$^\circ$).
  $V_{\Xi}=$ $-14$ MeV and $v^0_{\Xi N,\Lambda\Lambda}=$ 500 MeV
  were used.
  The labels $0^+(s^2_\Lambda)$, $1^-(s_\Lambda p_\Lambda)$ and 
  $2^+(p^2_\Lambda)$ denote the $J^\pi$ $\Lambda\Lambda$ nuclear 
  states of $(0s_\Lambda)^2$, $(0s_\Lambda)(0p_\Lambda)$ and $(0p_\Lambda)^2$
  coupled with $^{14}{\rm C}(0^+)$, respectively. 
  The labels $2^+(s^2_\Lambda)$, $1^-(2^+ \otimes s_\Lambda p_\Lambda)$ and 
  $2^+(2^+ \otimes p^2_\Lambda)$ denote the states of 
  $(0s_\Lambda)^2$, $(0s_\Lambda)(0p_\Lambda)$ and $(0p_\Lambda)^2$ coupled with 
  $^{14}{\rm C}(2^+)$, respectively. 
  }
  \end{center}
\end{figure}

In Fig.~\ref{fig:2}(b), we show the calculated spectra 
with the $\Lambda\Lambda $--$\Xi$ coupling potential 
when $V_{\Xi}=$ $-14$ MeV.
We recognize that the shape of these spectra 
is quite sensitive to the value of $v^0_{\Xi N,\Lambda\Lambda}$, 
and it is obvious that no $\Xi N$--$\Lambda\Lambda$ 
coupling cannot describe the spectrum of the $\Lambda\Lambda$ states 
below the ${^{14}{\rm C}}+\Lambda+\Lambda$ threshold.
The calculated spectrum for $v^0_{\Xi N,\Lambda\Lambda}=$ $500$ MeV has 
a fine structure of the $\Lambda\Lambda$ excited states 
in $^{\,16}_{\Lambda\Lambda}{\rm C}$.
We find that significant peaks of the $1^-$ excited states with 
${^{14}{\rm C}{(0^+)}}\otimes s_{\Lambda}p_{\Lambda}$ 
at $\omega=$ 362.1 MeV ($B_{\Lambda\Lambda} =$ 15.1 MeV)
and 
${^{14}{\rm C}{(2^+)}}\otimes s_{\Lambda}p_{\Lambda}$ 
at $\omega=$ 368.5 MeV ($B_{\Lambda\Lambda} =$ 8.7 MeV),
and small peaks of the $2^+$ excited states 
with ${^{14}{\rm C}(0^+)} \otimes p^2_\Lambda$
at $\omega=$ 373.8 MeV ($B_{\Lambda\Lambda} =$ 3.4 MeV)
and ${^{14}{\rm C}(2^+)} \otimes p^2_\Lambda$
at $\omega=$ 380.4 MeV ($B_{\Lambda\Lambda} =$ $-3.2$ MeV). 
This result arises from the fact that the high momentum transfer 
$q_{\Xi^-}\simeq$ 400 MeV/c leads to a preferential population of 
the spin-stretched $\Xi^-$ doorways states 
followed by the 
$
[{^{15}{\rm N}(1/2^-,3/2^-)}\otimes s_{\Xi^-}]_{1^-}
\to
[{^{14}{\rm C}(0^+,2^+)}\otimes s_{\Lambda}p_{\Lambda}]_{1^-}
$
and
$
[{^{15}{\rm N}(1/2^-,3/2^-)}\otimes p_{\Xi^-}]_{2^+}
\to
[{^{14}{\rm C}(0^+,2^+)}\otimes p^2_{\Lambda}]_{2^+}
$ transitions,
to which a sum of their continuum states may contribute predominately 
in the ($K^-$, $K^+$) reactions.
Figure~\ref{fig:3} also displays partial-wave decomposition of 
the calculated inclusive spectrum for
$^{\,16}_{\Lambda\Lambda}$C in the $\Lambda\Lambda$ bound region
when $V_{\Xi}=$ $-14$ MeV and 
$v^0_{\Xi N,\Lambda\Lambda}=$ 500 MeV. 
The integrated cross sections at $\theta_{\rm lab}=$ $0^\circ$ 
for the $1^-$ excited states with 
${^{14}{\rm C}{(0^+)}} \otimes s_{\Lambda}p_{\Lambda}$
and 
${^{14}{\rm C}{(2^+)}} \otimes s_{\Lambda}p_{\Lambda}$
are respectively
\begin{eqnarray}
{d\sigma \over d\Omega_L}\big[
{^{~16}_{\Lambda\Lambda}{\rm C}}(1^-)\big]
\simeq \ 7 \ \mbox{nb/sr and} \ 12 \ \mbox{nb/sr},  
\end{eqnarray}
where the $\Xi^-$ admixture probabilities of these states amount 
to $P_{\Xi^-}=$ 5.2\% and 8.8\%, respectively.
It should be noticed that the cross sections are on 
the same order of magnitude as those for the 
${1^-}$ and ${2^+}$ quasibound states that are located at 
$B_{\Xi^-}=$ 6.8 MeV and 0.5 MeV, respectively,
in the $^{16}_{\Xi^-}$C hypernucleus.
Therefore, such $\Lambda\Lambda$ excited states 
below the ${^{14}{\rm C}}+\Lambda+\Lambda$ threshold
will be measured experimentally at the J-PARC 
facilities \cite{NagaeE05}.

\Figuretable{FIG. 3}

On the other hand, it is extremely difficult to populate
the $0^+$ ground state with 
${^{14}{\rm C}{(0^+)}}\otimes s_{\Lambda}^2$ 
at $\omega \simeq$ 352.3 MeV ($B_{\Lambda\Lambda} \simeq$ 24.9 MeV)
and also the $2^+$ excited state with 
${^{14}{\rm C}{(2^+)}}\otimes s_{\Lambda}^2$ 
at $\omega \simeq$ 359.6 MeV ($B_{\Lambda\Lambda} \simeq$ 17.5 MeV)
in the one-step mechanism via $\Xi^-$ doorways 
in the ($K^-$,~$K^+$) reactions. 
The high momentum transfer of ${q_\Xi} \simeq$ 400 MeV/c
necessarily leads to the non-observability with $\Delta L=$ 0.
Thus the integrated cross section of the $0^+$ state is found 
to be about 0.02 nb/sr, 
of which the $q$ dependence is approximately governed 
by a factor of ${\rm exp}(-{1 \over 2}(\tilde{b}q_{\Xi})^2)$ 
where a size parameter $\tilde{b}=$ 1.84 fm.
There is no production in the $2^+$ state with 
${^{14}{\rm C}{(2^+)}}\otimes s_{\Lambda}^2$ 
under the angular-momentum conservation in the $^{16}{\rm O}(K^-, K^+)$ 
reactions by the one-step mechanism. 
The contribution of these states to the $\Lambda\Lambda$ spectrum in 
the one-step mechanism is completely different from that in 
the two-step mechanism 
as obtained in Refs.~\cite{Dover80,Baltz83}. 

In the ($K^-$, $K^+$) reaction, 
$\Lambda\Lambda$ hypernuclear states can be also populated 
by the two-step mechanism, $K^-p$ $\to$ $\pi^0\Lambda$ followed by 
$\pi^0p$ $\to$ $K^+\Lambda$ \cite{Dover80,Baltz83,Iijima92}, 
as shown in Fig.~\ref{fig:1}(a).
Following the procedure by Dover \cite{Dover80,Iijima92}, 
a crude estimate can be obtained for the contribution of this
two-step processes in the eikonal 
approximation using a harmonic oscillator model.
The cross section at $0^\circ$ for quasielastic $\Lambda\Lambda$ 
production at $p_{K^-}=$ 1.8 GeV/c in the two-step mechanism, 
which is summed over all final state, is given \cite{Iijima92} as
\begin{eqnarray}
\sum_f \Big({d\sigma_f^{(2)} \over d\Omega_L}\Big)_{0^\circ}
& \approx & {{2 \pi \xi } \over p^2_{\pi}}
\Big\langle {1 \over r^2} \Big\rangle
\Big(\alpha {d\sigma 
\over d\Omega_L} \Big)_{0^\circ}^{K^-p \to \pi^0 \Lambda} \nonumber\\
&& \times \Big(\alpha {d\sigma 
\over d\Omega_L} \Big)_{0^\circ}^{\pi^0p \to K^+ \Lambda} N_{\rm eff}^{pp},
\end{eqnarray}
where $\xi =$ 0.022--0.019 mb$^{-1}$ is a constant nature of the 
angular distributions of the two elementary processes, 
$p_\pi\simeq$ 1.68 GeV/c is the intermediate pion momentum, and 
$\langle 1/r^2 \rangle\simeq$ 0.028 mb$^{-1}$ is the mean inverse-square 
radial separation of the proton pair. 
$N_{\rm eff}^{pp}\simeq$ 1 is the effective number of proton pairs 
including the nuclear distortion effects \cite{Dover80}.
The elementary laboratory cross section 
$(\alpha {d\sigma/d\Omega_L})_{0^\circ}$ is estimated to be 1.57--1.26 mb/sr 
for ${K^-p \to \pi^0 \Lambda}$ and 0.070--0.067 mb/sr 
for ${\pi^0p \to K^+ \Lambda}$ depending on the nuclear medium corrections. 
This yields 
\begin{equation}
\sum_f \Big({d\sigma_f^{(2)} \over d\Omega_L}\Big)_{0^\circ} 
\simeq 0.06\mbox{--}0.04 \ \mbox{$\mu$b/sr}, 
\end{equation}
which is half smaller than $\sim$0.14 $\mu$b/sr at 1.1 GeV/c. 
Considering a high momentum transfer $q \simeq$ 400 MeV/c 
in the ($K^-$, $K^+$) reactions, 
we expect that the production probability for the $\Lambda\Lambda$ bound 
states does not exceed  1\% in the quasielastic $\Lambda\Lambda$ 
production, 
so that an estimate of the $\Lambda \Lambda$ hypernucleus 
in the two-step mechanism may be on the order of 0.1--1 nb/sr. 
This cross section is smaller than the cross section for the 
$\Lambda\Lambda$ $1^-$ states we mentioned above 
in the one-step mechanism.
Consequently, we believe that the one-step mechanism 
acts in a dominant process in the ($K^-$,~$K^+$) reaction 
at 1.8 GeV/c ($0^\circ$) 
when $v^0_{\Xi N,\Lambda\Lambda}=$ 400--600 MeV. 
This implies that the ($K^-$, $K^+$) spectrum provides 
valuable information concerning $\Xi N$--$\Lambda\Lambda$ 
dynamics in the $S=$ $-$2 systems 
such as $\Lambda\Lambda$ and $\Xi$ hypernuclei,
which are often discussed in a full coupling scheme \cite{Nemura05}.

\section{Summary and Conclusion}

We have examined theoretically production of doubly strange hypernuclei 
in the DCX $^{16}$O($K^-$,~$K^+$) reaction at 1.8 GeV/c  
within DWIA calculations using coupled-channel Green's functions.  
We have shown that the $\Xi^-$ admixture in the $\Lambda\Lambda$ hypernuclei
plays an essential role in producing the $\Lambda\Lambda$ states 
in the ($K^-$,~$K^+$) reaction. 

In conclusion, 
the calculated spectrum for the $^{~16}_{~\Xi^-}$C and 
$^{~16}_{\Lambda\Lambda}$C hypernuclei 
in the one-step mechanism $K^-p \to K^+\Xi^-$ via $\Xi^-$ 
doorways predicts promising peaks of the $\Lambda\Lambda$ bound 
and excited states in the $^{16}$O($K^-$,~$K^+$) reactions 
at 1.8 GeV/c (0$^\circ$). 
It has been shown that 
the integrated cross sections 
for the significant $1^-$ excited states 
in $^{~16}_{\Lambda\Lambda}$C are on the order of 7--12 nb/sr 
depending on the $\Xi N$--$\Lambda\Lambda$ coupling strength and 
also the attraction in the $\Xi$-nucleus potential. 
The $\Xi^-$ admixture probabilities are on the order of 5--9\%.
The sensitivity to the potential parameters indicates 
that the nuclear ($K^-$,~$K^+$) reactions 
have a high ability for the theoretical analysis of 
precise wave functions in the $\Lambda\Lambda$ and $\Xi$ hypernuclei.
New information on $\Lambda\Lambda$--$\Xi$ 
dynamics in nuclei from the ($K^-$,$K^+$) data at J-PARC 
facilities \cite{NagaeE05}
will bring the $S=-2$ world development in nuclear physics.

\begin{acknowledgments}
The authors are obliged to T. Fukuda, Y. Akaishi, D.E. Lanskoy, 
T. Motoba and T. Nagae for many discussions.
This work was supported by Grants-in-Aid for Scientific Research on
Priority Areas (Nos.~17070002 and 20028010) and 
for Scientific Research (C) (No.~225402294).
\end{acknowledgments}


\clearpage

\end{document}